\pdfoutput=1
\RequirePackage{ifpdf}
\ifpdf 
\documentclass[pdftex]{sigma}
\else
\documentclass{sigma}
\fi

\def\onehalf{{\textstyle \frac12}}
\def\ii{{\rm i}}
\def\dd{{\rm d}}
\def\MM{{\bf M}}
\def\ZZ{{\sf Z}}
\def\LCT{{\cal C}}

\def\totder#1#2{\frac{\dd #1}{\dd #2}}

\def\ssr#1{{\scriptscriptstyle\rm #1}}
\def\tsty#1#2{{\textstyle\frac{#1}{#2}}}
\def\trans{{\scriptscriptstyle\top}}
\def\real{{\sf R}}

\def\Lie#1{\hbox{\sf #1}}
\def\Poissbra#1#2{{}\{ #1 , #2 \}}
\def\matdos#1#2#3#4{\left(\begin{matrix}\displaystyle{#1}&
			\displaystyle{#2}{}_{\mathstrut}
		\cr \displaystyle{#3}&\displaystyle{#4}\cr \end{matrix}\right)}
\def\matricita#1#2#3#4{{\textstyle\Big({#1\atop #3}\ {#2\atop #4}\Big)}}

\def\vecdos#1#2{\left( \begin{matrix}\displaystyle{#1}{}_{\mathstrut}\cr
		\displaystyle{#2}\cr\end{matrix}\right)}
\def\vectorcito#1#2{{\textstyle\Big({#1\atop #2}\Big)}}

\newcommand{\be}{\begin{equation}}
\newcommand{\ee}{\end{equation}}
\newcommand{\bea}{\begin{eqnarray}}
\newcommand{\eea}{\end{eqnarray}}

\begin{document}

\allowdisplaybreaks

\renewcommand{\thefootnote}{$\star$}

\renewcommand{\PaperNumber}{033}

\FirstPageHeading

\ShortArticleName{A Top-Down Account of Linear Canonical Transforms}

\ArticleName{A Top-Down Account of Linear Canonical Transforms\footnote{This
paper is a contribution to the Special Issue ``Superintegrability, Exact Solvability, and Special Functions''. The full collection is available at \href{http://www.emis.de/journals/SIGMA/SESSF2012.html}{http://www.emis.de/journals/SIGMA/SESSF2012.html}}}

\Author{Kurt Bernardo WOLF}

\AuthorNameForHeading{K.B.~Wolf}

\Address{Instituto de Ciencias F\'{\i}sicas,
		 Universidad Nacional Aut\'onoma de M\'exico,\\
Av.\ Universidad s/n, Cuernavaca, Mor.\ 62210, M\'exico}
\Email{\href{mailto:bwolf@fis.unam.mx}{bwolf@fis.unam.mx}}
\URLaddress{\url{http://www.fis.unam.mx/~bwolf/}}

\ArticleDates{Received April 24, 2012, in f\/inal form June 01, 2012; Published online June 06, 2012}

\Abstract{We contend that what are called Linear Canonical Transforms (LCTs)
should be seen as a part of the theory of unitary irreducible
representations of the `$2{+}1$' Lorentz group. The integral kernel
representation found by Collins, Moshinsky and Quesne, and the radial
and hyperbolic LCTs introduced thereafter, belong to the discrete and
continuous representation series of the Lorentz group in its
parabolic subgroup reduction. The reduction by the elliptic and
hyperbolic subgroups can also be considered to yield LCTs that act on
functions, discrete or continuous in other Hilbert spaces. We gather the
summation and integration kernels reported by Basu and Wolf when
studiying all discrete, continuous, and mixed representations of the
linear group of $2\times2$ real matrices. We add some comments on why
all should be considered canonical.}

\Keywords{linear transforms; canonical transforms; Lie group \Lie{Sp$(2,\real)$}}

\Classification{20C10; 20C35; 33C15; 33C45}

\renewcommand{\thefootnote}{\arabic{footnote}}
\setcounter{footnote}{0}

\section{Introduction}   \label{sec:one}

Linear canonical transforms (LCTs) have been developed as the
dif\/fraction integral kernel for generic paraxial optical systems by
Stuart A.~Collins \cite{Collins}, and also def\/ined as the group of
unitary integral transforms that preserves the basic Heisenberg
uncertainty relation of quantum mechanics in $D=1$ or higher
dimensions by Marcos Moshinsky and Christiane Quesne
\mbox{\cite{Moshinsky-Quesne-can1,Moshinsky-Quesne-osc,Moshinsky-Quesne-can2}}.
Further, LCTs
can be seen as the group actions generated by the Lie algebra of
quadratic Hamiltonian operators \cite{CT-I}. Also, radial
\cite{MM-THS-KBW,CT-II} and hyperbolic \cite{CT-IV} canonical
transforms have been def\/ined after separation of variables in $D\ge2$
dimensions. More recently there has been interest in canonical
transformations that are represented by matrices, inf\/inite,
semi-inf\/inite or f\/inite-dimensional~-- the latter due to the f\/inite
capacity of measuring and storing devices, but with serious
concomitant dif\/f\/iculties.

In both the optical Lagrangian formulation \cite{Collins} or in
the quant\-um-mech\-anic\-al Hamiltonian approach
\cite{Moshinsky-Quesne-can1},
one arrives at a parametrization of all possible linear
transformations through a $2D\times2D$ real symplectic matrix
$\MM\in\Lie{Sp($2D,\real$)}$. For reference we gather here the
basic $D=1$-dimensional formulas. We consider the matrix
\begin{gather}
	\MM=\begin{pmatrix} a& b \\ c& d\end{pmatrix} , \qquad ad-bc=1,
				\label{Mabcd}
\end{gather}
as the presentation of the group \Lie{Sp($2,\real$)}, which is equal to
the groups of all real $2\times2$ matrices of unit determinant
\Lie{SL($2,\real$)}, and is isomorphic through a complex similarity
transformation to the group \Lie{SU($1,1$)} of $2\times2$ pseudo-unitary
matrices. This group of $2\times2$ matrices in fact covers twice, i.e.,
is 2:1 homomorphic to the `$2{+}1$'-Lorentz group of
$3\times3$ pseudo-orthogonal matrices ${\bf L}(\pm\MM)\in \Lie{SO($2,1$)}$,
\begin{gather}
		\begin{pmatrix} a& b \\ c& d\end{pmatrix}
\ \overset{2:1}{\longleftrightarrow} \
\begin{pmatrix}
\onehalf(a^2{-}b^2{-}c^2{+}d^2) & bd{-}ac & \onehalf(a^2{-}b^2{+}c^2{-}d^2)\\
cd{-}ab & ad{+}bc & -cd{-}ab\\
\onehalf(a^2{+}b^2{-}c^2{-}d^2) & -bd{-}ac & \onehalf(a^2{+}b^2{+}c^2{+}d^2)
\end{pmatrix}.
				\label{sp2R-so21}
\end{gather}
These accidental iso- and homomorphisms between symplectic and
relativity groups have been exploited to describe and relate various
physical and optical models \cite{Han,Monzon,Mukunda,Yonte}.

However, in spite of its apparent simplicity, the group of
$2\times2$ matrices \eqref{Mabcd} has an inf\/inite cover group
$\overline{\Lie{Sp}}\Lie{($2,\real$)}$. It is the double cover~-- the {\it metaplectic\/} group \Lie{Mp($2,\real$)}~-- which is
represented by the well-known {\it integral form\/} of linear
canonical transforms ${\LCT}_\ssr{M}\equiv {\LCT}\!\matricita
abcd$ of the usual Hilbert space of functions
$f(x)\in{\cal L}^2(\real)$,
\begin{gather}
	f_\ssr{\!M}(x)  \equiv  (\LCT_\ssr{M} f)(x)
	= \int_{\real}\dd x'\,C_\ssr{M}(x,x') f(x'),
				\nonumber\\ 
	C_\ssr{M}(x,x')  :=  \frac{1}{\sqrt{2\pi\ii b}}
		\exp\left(\frac{\ii}{2b}\big(d x^2-2xx'+a x^{\prime\,2}\big)\right),
						\label{integral-C1}
\end{gather}
where the phase of the prefactor is taken to be
\begin{gather*}
	\frac1{\sqrt{2\pi\ii b}}=\frac1{\sqrt{2\pi |b|}}
			\exp\left(-\ii\tsty14\pi\,{\rm sign}\,b\right).
\end{gather*}
Unitarity is evident in that $C_{\ssr{M}^{-1}}(x,x') =
C_\ssr{M}(x',x)^*$.

Phases are quite delicate here; Collins did not consider the group
property of the transforms~\cite{Collins}; yet Moshinsky and Quesne
\cite{Moshinsky-Quesne-can1} realized that
$\LCT_{\ssr{M}_1}\LCT_{\ssr{M}_2}=\pm\LCT_{\ssr{M}_1\ssr{M}_2}$, with
a sign depending on the signs of the $b_1$, $b_2$ and $b_{12}$
elements in a rather complicated way \cite[Chapter~9]{KBW-book}.  The
double-cover issue can be simplif\/ied by observing that for ${\bf
F}=\matricita0{1}{-1}0$, the kernel~\eqref{integral-C1} is that of the
Fourier transform ${\cal F}$, but for a phase: $\LCT_\ssr{F}=
e^{-\ii\pi/4}{\cal F}$. Since ${\cal F}^4={\it1}$, only
$\LCT_\ssr{F}^8$ will return the cycle to the unit~${\it 1}$. In the
limit $b\to0$ from the lower complex half-plane, the kernel~\eqref{sp2R-so21} becomes a Dirac delta,
\begin{gather}
	\lim_{b\to0}C_\ssr{M}(x,x') = \frac{\exp(\ii c x^2/2a)}{\sqrt{a}}
			 \delta\left(x'-\frac xa\right).
					\label{limbto0}
\end{gather}
We may skip further detailed consideration of the $D=1$ integral
linear canonical transforms and their properties, which are mostly
standard knowledge, and from whose $D=2$ case one can build
the radial and hyperbolic LCTs in a bottom-up construction.
Rather, the purpose of this review is to give a top-down panorama of LCTs.

In Section~\ref{sec:two} we return to the Lie algebra
\Lie{so($2,1$)} of the Lorentz group \Lie{SO($2,1$)} realized by
second-order dif\/ferential operators corresponding to harmonic and
repusive oscillators, with a~singular centrifugal or centripetal
potential, and the generator of scaling. Linear combinations yield
free propagation and a square-radius coordinate. In Section~\ref{sec:three} we list the eigenfunctions and spectra of those
operators for the centrifugal case that fall into the Bargmann ${\cal
D}_k$ discrete series of representations. There are three subgroup
orbits that are examined in three subsections: elliptic, parabolic,
and hyperbolic. Section~\ref{sec:four} follows the same structure for
the centripetal case that fall into the Bargmann ${\cal
C}^\varepsilon_s$ continuous series. This is what we deem to be
six faces that \Lie{Sp($2,\real$}) linear canonical transforms can
show in various Hilbert spaces. The resulting matrix and integral
kernels were computed in 1981 by Basu and Wolf~\cite{Basu-KBW}, so
the results are not new, but are gathered here for the f\/irst time as
proper, unitary LCTs. The concluding Section~\ref{sec:five}
of\/fers some reasons to call all representations canonical, and some
comments on f\/inite-dimensional approximations to these transforms.

\section{The Lorentz algebra}   \label{sec:two}

The generic form of the second-order dif\/ferential operator
realization of the Lorentz Lie algebra \Lie{so($2,1$)} can be
obtained from the oscillator algebra~\cite{Moshinsky-Quesne-osc},
adding a term $\gamma/r^2$ to the second derivative, whose
interpretation is that of a centrifugal potential as
$\gamma=\overline{m}^2-\frac14\ge-\frac14$ for solutions of angular
momentum $\overline{m}\in\ZZ$. Separation into hyperbolic coordinates
yields $\gamma<-\frac14$ when added by a centripetal
force. The formal operators are associated with Hamiltonians, and
are generators of \Lie{Sp($2,\real$)} subgroups as follows:
\begin{alignat}{3}
&	J^\gamma_0  :=
		\frac14\left({-\totder{^2}{r^2}}+\frac\gamma{r^2}
		+r^2\right),\qquad && \exp(\ii\phi J^\gamma_0)\mapsto
		\matdos{\cos\onehalf\phi}{-\sin\onehalf\phi
			}{\sin\onehalf\phi}{\phantom{-}\cos\onehalf\phi},&\nonumber\\
& J^\gamma_1  :=
		\frac14\left({-\totder{^2}{r^2}}+\frac\gamma{r^2}
		-r^2\right), \qquad &&  \exp(\ii\zeta J^\gamma_1)\mapsto
		\matdos{\phantom{-}\cosh\onehalf\zeta}{-\sinh\onehalf\zeta
			}{-\sinh\onehalf\zeta}{\phantom{-}\cosh\onehalf\zeta},& \label{J2andM}\\
&	J^\gamma_2:=     -\frac{\ii}2\left(r\totder{}{r}+\frac12
		   \right), \qquad &&  \exp(\ii\alpha J^\gamma_2)\mapsto
		\matdos{\exp {-\onehalf\alpha}}{0}{0}{\exp {+\onehalf\alpha}},\nonumber
\end{alignat}
where the `harmonic oscillator' $J^\gamma_0$ generates the compact
({\it elliptic\/}) \Lie{SO($2$)} subgroup (and its co\-vers), while
the `repulsive oscillator' $J^\gamma_1$ and `scaling operator'
$J^\gamma_2$ generate equivalent noncompact ({\it hyperbolic\/})
subgroups \Lie{SO($1,1$)}. Their commutation relations are
\begin{gather}
	[J^\gamma_1,J^\gamma_2]=-\ii J^\gamma_0,\qquad
	[J^\gamma_2,J^\gamma_0]=\ii J^\gamma_1,\qquad
	[J^\gamma_0,J^\gamma_1]=\ii J^\gamma_2.
			\label{com-rel-so21}
\end{gather}

Also relevant are the linear combinations
\begin{alignat}{3}
&	J^\gamma_+  :=   J^\gamma_0+J^\gamma_1
				=\frac12\left({-\totder{^2}{r^2}}+\frac\gamma{r^2}
		    \right), \qquad &&  \exp(\ii b J^\gamma_+)\mapsto
		\matdos{1}{-b}{0}{1},&\nonumber\\
&	J^\gamma_-  :=  J^\gamma_0-J^\gamma_1
			= \onehalf r^2, \qquad &&
			\exp(\ii c J^\gamma_-)\mapsto \matdos{1}{0}{c}{1},&
				      \label{JMinusandM}
\end{alignat}
where the `free system' $J^\gamma_+$ and the `square-radius'
$J^\gamma_-$ generate the Euclidean ({\it parabolic\/})
subgroups \Lie{ISO($1$)}, which are equivalent under the Fourier
transform. Note carefully that these are {\it not\/} `raising and
lowering' operators for discrete eigenvector bases. Instead and
particularly, $J^\gamma_-$~determines the {\it diagonal radial
position\/} operator with respect to which we refer the
eigenfunctions of all other operators, as done below. These
eigenfunctions depend crucially on the value of~$\gamma$ in~\eqref{J2andM}, as given by the Casimir eigenvalue
\begin{gather}
	C := (J^\gamma_1)^2+(J^\gamma_2)^2-(J^\gamma_0)^2 = \kappa{\it1},
			\nonumber\\ 
	\kappa  =  -\tsty14\gamma+\tsty3{16} =: k(1-k),\qquad
		k=\onehalf\left(1\pm\sqrt{\gamma+\tsty14}\right),
			\label{Bargmann-index}
\end{gather}
where $k$ is the {\it Bargmann index\/} \cite{Bargmann-1947} that
determines (up to
parities) the essentially self-adjoint irreducible representations of
the algebra in a Hilbert space ${\cal L}^2(\real^+)$ with measure~$\dd r$~-- which are unexpectedly imbricate.

\section[The ${\cal D}_k$ canonical transforms]{The $\boldsymbol{{\cal D}_k}$ canonical transforms}   \label{sec:three}

We consider f\/irst the case when the coef\/f\/icient $\gamma$ is of
centrifugal origin in $D=2$ using polar coordinates $x=r\cos\theta$
and $y=r\sin\theta$, $r\in\real^+$, $\theta$ modulo $2\pi$. We want
the measure to be $\dd r$, so after similarity transformation,
$\sqrt{r}\nabla^2/\sqrt{r}=\partial_r^2 +
r^{-2}(\frac14+\partial^2_\theta)$, with
$\partial_\theta^2\mapsto{-\overline{m}^2}$ for the subspaces of
angular momentum $\overline{m}\in\ZZ$ so that
$k=\onehalf(|\overline{m}|+1)\in\{\onehalf,1,\frac32,\ldots\}$. This
is the Bargmann {\it discrete\/} representation series ${\cal
D}_k^\pm$ \cite{Bargmann-1947}.\footnote{The same were named by I.M.~Gel'fand and M.A.~Na\u{\i}mark to be the {\it complementary\/}
representation series~\mbox{\cite{Gelfand-Naimark,Naimark}}. Note that
$k\leftrightarrow 1-k$ correspond to the same $\kappa$; the interval
$0<k<1$ is {\it exceptional\/} in that the operators have a
one-parameter family of self-adjoint extensions~\cite{Fortschritte},
and also harbor the exceptional, or {\it supplementary\/} irreducible
representation series~\cite{Bargmann-1947,Gelfand-Naimark,Naimark}.
We shall be working within the Friedrichs extension.} The
eigenfunctions and spectra of the f\/ive operators \eqref{J2andM} and
\eqref{JMinusandM}, are \cite{Basu-KBW}
\begin{gather}
	{}^0\Phi^k_m(r) =
		\sqrt{\frac{2\,n!}{(2k{-}n{-}1)!}}
			 r^{2k-1/2}e^{-r^2/2}\,L_n^{(2k-1)}(r^2)
		\nonumber \\
\hphantom{{}^0\Phi^k_m(r)}{} =   \frac{\sqrt{2\,(2k{+}n{-}1)!}}{\Gamma(2k) \sqrt{n!}}
		 r^{2k-1/2}e^{-r^2/2}\,{}_1\!F_{\!1}\left({-n\atop2k}; r^2\right), \qquad m=k{+}n,\quad  n\in\ZZ,\nonumber
		 	\\
	{}^{+}\Phi^{k}_\rho(r)  =  e^{\ii\pi k}\sqrt{\rho r}
					J_{2k-1}(\rho r),\qquad \rho\in\real^+,\nonumber\\
	{}^{-}\Phi^{k}_\rho(r)  =  \delta(\rho-r),\qquad \rho\in\real^+,\nonumber\\
	{}^{1}\Phi^{k}_\mu(r)  =
		\frac{e^{\ii\pi(2k+\mu)/2}\,2^{\ii\mu}
		\Gamma(k{+}\ii\mu)}{\Gamma(2k) \sqrt{\pi}} \frac1{\sqrt{r}}
			M_{\ii\mu,k-1/2}(-\ii r^2), \nonumber \\
\hphantom{{}^{1}\Phi^{k}_\mu(r)}{} =  \frac{e^{\ii\pi(k+\mu)/2} 2^{\ii\mu}
		\Gamma(k{+}\ii\mu)}{\Gamma(2k) \sqrt{\pi}}
			r^{2k-1/2}e^{\ii r^2/2}\,
			{}_1\!F_{\!1}\left({k{-}\ii\mu\atop2k}; -\ii r^2\right),\qquad \mu\in\real,
						\nonumber\\
	{}^{2}\Phi^{k}_\mu(r)  =
			\frac1{\sqrt{\pi}} r^{-1/2+\ii\mu},
							\qquad \mu\in\real,
									  \label{Phiminus}
\end{gather}
where $L_n^{(\mu)}$ are the Laguerre polynomials, $J_\mu$ is
the Bessel function of the f\/irst kind, and~$M_{\lambda,\mu}$
is one of the Whittaker functions.
In Dirac notation, ${}^X\!\Phi^k_\lambda(r)\equiv
{}_-\!\langle k,r|k,\lambda\rangle_{\!X}$, with $X\in\{0,1,2,+,-\}$
indicating the eigenkets of $J^\gamma_{\!X}$, with eigenvalues
$\lambda\in\Sigma_X$ in the spectra of~\eqref{Phiminus}.

Next, we should obtain the transformation of the functions in
\eqref{Phiminus} under the generic LCT
$\LCT_\ssr{M}\equiv\LCT\matricita abcd$. To this
end for each generator $J^\gamma_{\!X}$ in the lists \eqref{J2andM}
and \eqref{JMinusandM} we decompose $\LCT_\ssr{M}$ into
a right-ordered product of the subgroup with a diagonal-plus-phase
transformation as \eqref{limbto0},
\begin{gather}
	\LCT\matricita abcd = \LCT\matricita{a'}0{c'}{1/a'}
				 \exp\big(\ii\alpha J^\gamma_{\!X}\big),		\label{Iwasawa-type}
\end{gather}
where $a'$, $c'$ and $\alpha$ will be algebraic and trigonometric
functions of $a$, $b$, $c$, $d$, respecting $ad-bc=1$.
The right factor applied to ${}^X\!\Phi^k_\lambda(r)$ will
multiply it by the phase $e^{\ii\lambda\alpha}$, while the left
factor will be given by \eqref{limbto0}, with the sole precaution of
taking $1/\sqrt{a} \equiv (\hbox{sign}\,a)^{2k}/\sqrt{|a|}$
stemming from the radial reduction of the $D=2$ case with angular
momentum.

\subsection[Elliptic basis: lower-bound discrete ${\cal D}_k$-LCTs]{Elliptic basis: lower-bound discrete $\boldsymbol{{\cal D}_k}$-LCTs}

We consider f\/irst the eigenbasis  of the compact generator
$J^\gamma_0$, ${}^0\Phi^k_m(r)$, in the ${\cal D}^+_k$ Bargmann
representation series, whose eigenvalues are lower-bound
and discrete, $\lambda\equiv m=k+n$, $n\in\ZZ^+_0$.  The decomposition
\eqref{Iwasawa-type} yields $a'=\sqrt{a^2+b^2}$, $a'
c'=ac+bd$ and $e^{\ii\alpha}=(a{-}\ii b)/(a{+}\ii b)$. Then,
\begin{gather}
	\LCT\matricita abcd\,{}^0\!\Phi^k_m(r)
		=\left(\frac{a{-}\ii b}{a{+}\ii b}\right)^m
		\frac{\exp\left(\ii r^2\frac{ac{+}bd}{2(a^2{+}b^2)}\right)
				}{(a^2+b^2)^{1/4}} \,
		{}^0\!\Phi^k_m\left(\frac{r}{\sqrt{a^2{+}b^2}}\right).
				\label{LCT-on-Phi0}
\end{gather}
From this we can f\/ind the representation matrices of the group,
\begin{gather*}
	{}^X\!\!D^k_{\lambda,\lambda'}(\MM)= \int_0^\infty {}_X\!\langle k,\lambda
		|k,r\rangle_{\!-}\,\dd r\,{}_-\!\langle k,r|{\LCT_\ssr{M}}
		|k,\lambda'\rangle_{\!X}, 
\end{gather*}
by straightforward integration of the conf\/luent hypergeometric
functions that appear in  the tables of Gradshte\u{\i}n and Ryzhik~\cite{GR}.
In the \Lie{SO($2$)} basis, this is a matrix with rows and columns
numbered by $m=k+n$, $n\in\ZZ_0^+$, that was found by Bargmann
\cite{Bargmann-1947},
\begin{gather}
	{}^0\!D^k_{m,m'}\matricita abcd
	:=\left({}^0\!\Phi^k_m,\,\LCT\matricita abcd\,{}^0\!\Phi^k_{m'}\right)
	=\int_0^\infty\dd r\, {}^0\!\Phi^k_m(r)^*\,
		\LCT\matricita abcd\,{}^0\!\Phi^k_{m'}(r) \label{DoC00}\\
\hphantom{{}^0\!D^k_{m,m'}\matricita abcd}{}
		 = \frac{2^{2k}\Gamma(m{+}m')}{
		  \sqrt{\Gamma(k{+}m) \Gamma(1{-}k{+}m)
				\Gamma(k{+}m') \Gamma(1{-}k{+}m')}} \nonumber\\
\hphantom{{}^0\!D^k_{m,m'}\matricita abcd :=}  {}\times
			[(d{-}a){-}\ii(b{+}c)]^{m-k}
			[(a{-}d){-}\ii(b{+}c)]^{m'-k}
			[(a{+}d){+}\ii(b{-}c)]^{-m-m'}\nonumber \\
\hphantom{{}^0\!D^k_{m,m'}\matricita abcd :=}
{}\times {}_2\!F_{\!1}\left(
			{k{-}m,\ k{-}m'\atop 1{-}m{-}m'}\,;\,
			\frac{a^2{+}b^2{+}c^2{+}d^2{+}2}{a^2{+}b^2{+}c^2{+}d^2{-}2}
					\right).
						\label{D0-2}
\end{gather}
Then, when ${\bf f}\equiv\{f_n\}_{n=0}^\infty$ is a vector in the
Hilbert space of square-summable sequences $\ell^2(\ZZ_0^+)$, the
\Lie{Sp($2,\real$)} action
	\begin{gather}
	{\bf f}_\ssr{M} \equiv \LCT_\ssr{M}:{\bf f}
					= {}^0\!{\bf D}^k {\bf f},\qquad
	f_{\!\ssr{M};\,n}  \equiv  \displaystyle({\bf f}_\ssr{M})_n
		= \sum_{n'=0}^\infty
			{}^0\!\!D^k_{k+n,k+n'}({\bf M}) f_{n'},
				   \label{CT-0}
\end{gather}
is an LCT which is unitary in the Hilbert
space of sequences $\ell^2(\ZZ_0^+)$.

In fact, \eqref{D0-2} provides a unitary summation transform kernel for
every value $k>0$, not necessarily stemming from integer angular
momentum $\overline{m}$, where $k=\onehalf(|\overline{m}|+1) \in
\{\onehalf,1,\frac32,\ldots\}$ are single- and double-covers of the
\Lie{SO($2,1$)} group, i.e., single covers of \Lie{Sp($2,\real$)}.
When $k$ is quarter-integer (in particular $k=\frac14$ and $\frac34$, to be
revisited below), we have representations of the metaplectic group
\Lie{Mp($2,\real$)}, and various higher covers for fractional $k$'s.

In the ${\cal D}^+_k$ Bargmann representations, the generator
$J_0^\gamma$ has an equally-spaced spectrum $\{m\}_{m=k}^\infty$
which is bound from below by $k>0$. There is also a paired series of
representations ${\cal D}^-_k$ where that spectrum is upper-bound by
$-k$, i.e., $m=-k-n$, $n\in\ZZ^+_0$. This stems from the outer
algebra automorphism $J^\gamma_0\leftrightarrow-J^\gamma_0$
(reversing the sign of the spectrum),
$J^\gamma_1\leftrightarrow-J^\gamma_1$,
$J^\gamma_2\leftrightarrow J^\gamma_2$, and so
$J^\gamma_\pm\leftrightarrow-J^\gamma_\pm$. This generates an outer
group automorphism whose representations in any subgroup reduction
yields the ${\cal D}^-_k$ matrices or integral kernels
\begin{gather}
	{}^X\!\!D_{\lambda,\lambda'}^{k,(-)}\matricita abcd
		\leftrightarrow
	{}^X\!\!D_{\sigma\lambda,\sigma\lambda'}^{k,(+)}
		\matricita a{-b}{-c}d,
				\label{Dminus-reps}
\end{gather}
where $J_X^\gamma$ has spectrum $\{\lambda\}\in\Sigma_X$ and parity
$\sigma\in\{+,-\}$ under this automorphism. In fact, the matrices in~\eqref{Dminus-reps} are related by a similarity transformation with
$\matricita 100{-1}$, a non-symplectic matrix which describes
ref\/lection in geometric optics~\cite[Chapter~4]{GeomOpt}.

We have spoken of ${\cal D}^\pm_k$ for $k>0$. What about $k<0$? The
Bargmann index $k$ acts as a lower bound for the values of $m$ in
${\cal D}^+_k$, and as an upper bound in ${\cal D}^-_k$ in the
one-step recursion relations obtained with raising and lowering
operators on the eigenvectors of $J^\gamma_0$; but they are one-way
barriers. At $k=0$ the $m=\pm k$ lines cross, so that when $k<0$ is
integer or half-integer, the $m$'s have a lower bound at negative $k$
and an upper bound at positive $-k$. Raising and lowering the $m$'s
in this range will yield a $(2k{+}1)$-dimensional matrix, which is a
faithful, although {\it non\/}-unitary irreducible representation of
\Lie{Sp($2,\real$)}. Of course, a well-known theorem (see~\cite{Gilmore}) states that non-compact groups do not possess
f\/inite-dimensional unitary irreducible representations.

\subsection[Parabolic basis: continuous radial ${\cal D}_k$-LCTs]{Parabolic basis: continuous radial $\boldsymbol{{\cal D}_k}$-LCTs}

The eigenfunctions ${}^{+}\Phi^{k}_\rho(r)$ and
${}^{-}\Phi^{k}_\rho(r)$ in \eqref{Phiminus} of the parabolic generators
$J^\gamma_+$ and $J^\gamma_-$ in~\eqref{JMinusandM} form generalized
bases for ${\cal L}^2(\real^+)$. Consider thus the \Lie{Sp($2,\real$)}
representation given as an integral kernel by
\begin{gather}
	 {}^{-}\!D^k_{\rho,\rho'}\matricita abcd
	:=\left({}^-\!\Phi^k_\rho,
			\LCT\matricita abcd\,{}^-\!\Phi^k_{\rho'}\right)
	=\left({}^-\!\Phi^k_\rho, \LCT\matricita abcd
			\LCT\matricita 0{-1}10\,{}^+\!\Phi^k_{\rho'}\right)
								\label{DO-minmin1}\\
\hphantom{{}^{-}\!D^k_{\rho,\rho'}\matricita abcd}{}
= \left({}^-\!\Phi^k_\rho, \LCT\matricita b0d{1/b}
			\LCT\matricita 1{-a/b}01\,{}^+\!\Phi^k_{\rho'}\right)
								\nonumber\\
\hphantom{{}^{-}\!D^k_{\rho,\rho'}\matricita abcd}{}
	 = \frac{e^{-\ii\pi k}}{b} \sqrt{\rho\rho'}
		\exp\left(\ii\frac{d\rho^2{+}a\rho^{\prime\,2}}{2b}\right)
		 J_{2k-1}\left(\frac{\rho\rho'}{b}\right)
								\label{the-radial}\\
\hphantom{{}^{-}\!D^k_{\rho,\rho'}\matricita abcd}{}
 = \frac{2 (\rho\rho')^{2k-1/2}}{(2\ii b)^{2k}\,\Gamma(2k)}
		\exp\left(\ii\frac{d\rho^2{-}2\rho\rho'
							{+}a\rho^{\prime\,2}}{2b}\right)
			{}_1\!F_{\!1}\left(	{2k-\onehalf\atop 4k-1}\,;\,
					\frac{2\ii\rho\rho'}{b} \right).
								\label{DO-minmin2}
\end{gather}
This we recognize as the radial canonical transform kernel
\cite{MM-THS-KBW,CT-II} for angular momentum
$k=\onehalf(\overline{m}+1)$, which can be extended to $k>0$, that
acts unitarily on the functions ${\bf f}\equiv\{f(r)\}_{r\in\real^+}$
in the Hilbert space ${\cal L}^2(\real^+)$ with measure~$\dd r$. As
we had in \eqref{CT-0}, now
\begin{gather}
	{\bf f}_\ssr{M} \equiv \LCT_\ssr{M}:{\bf f}
					= {}^-\!{\bf D}^k({\bf M}) {\bf f},\qquad
	f_{\!\ssr{M}}(r)  \equiv   ({\bf f}_\ssr{M})(r)
		= \int_0^\infty \dd r'\,
			{}^-\!\!D^k_{r,r'}({\bf M}) f(r').
   \label{CT-rad}
\end{gather}

We regain the classical $D=1$ LCT kernel in \eqref{integral-C1} as the
direct sum of the representations $k=\frac14$ and $\frac34$, for
functions $f(x)$ whose domain is extended to $x\in\real$ through
writing them as the sum $f(x)= f^{(1/4)}_{\!e}(x)+f^{(3/4)}_{\!o}(x)$
with even and odd parity summands $f_{\!e}(-r):=f_{\!e}(r)$ and
$f_{\!o}(-r):=-f_{\!o}(r)$ respectively. For $2k{-}1=\mp1$, the
integral kernels contain $J_{-1/2}(z)=\sqrt{2/\pi z}\cos z$ and
$J_{+1/2}(z)=\sqrt{2/\pi z}\sin z$, $z=\rho\rho'/b$, with phases
$\ii^{2k}=e^{\ii\pi/4}$ and $\ii e^{\ii\pi/4}$. Their sum thus
yields $e^{\ii\pi/4}\,e^{-\ii xx'/b}$ in the oscillating Gaussian of
the original LCT kernel \eqref{integral-C1},
\begin{gather}
	 C_\ssr{M}(x,x')
	    =D^{(1/4)}_{\ssr{M}}(r,r') + D^{(3/4)}_{\ssr{M}}(r,r').
	 				\label{osc-rep-134}
\end{gather}

Had we chosen the eigenbasis of the free Hamiltonian $J^\gamma_+$
instead of the square-position $J^\gamma_-$, the Bessel function
${}^{+}\Phi^k_\rho(r)$ in \eqref{Phiminus} we would have the Hankel
transform of~\eqref{DO-minmin2} by ${\bf F}=\matricita0{1}{-1}0$, so
\begin{gather}
	{}^+\!D^k_{\rho,\rho'}\matricita abcd
	 	:=\left({}^+\!\Phi^k_\rho,
			\LCT\matricita abcd\,{}^+\!\Phi^k_{\rho'}\right)
	= {}^-\!D^k_{\rho,\rho'}\matricita{d}{-c}{-b}{a}.
				\label{Jplus-eigenbasis}
\end{gather}

\subsection[Hyperbolic basis: the face of ${\cal D}_k$-LCTs]{Hyperbolic basis: the face of $\boldsymbol{{\cal D}_k}$-LCTs}

There remain the eigenbases of the two equivalent noncompact
operators in the list~\eqref{J2andM}: the repulsive oscillator
Hamiltonian $J^\gamma_1$ and the
scaling generator $J^\gamma_2$. The latter is
the simpler of the two because its eigenfunctions,
${}^{2}\Phi^{k}_\mu(r)$ in \eqref{Phiminus}, are the Mellin transform
kernel with $\mu\in\real$ and $k>0$. Since we know the LCT action
$\LCT_\ssr{M}$ in the parabolic basis, \eqref{DO-minmin1}--\eqref{CT-rad},
we apply it to these eigenfunctions,
\begin{gather}
	\big(\LCT_\ssr{M}:{}^{2}\Phi^{k}_\mu\big)(r)
		 = \frac1{\sqrt{\pi}}\int_0^\infty\dd r'\,
			{}^-\!\!D^k_{r,r'}\matricita abcd
			 r^{\prime\,{-}1/2+\ii\mu}  \nonumber \\ 
\phantom{\big(\LCT_\ssr{M}:{}^{2}\Phi^{k}_\mu\big)(r)}{}		
 =  \frac{e^{-\ii\pi k}}{2^{k-\ii\mu}\sqrt{\pi}}
			\frac{\Gamma(k{+}\ii\mu)}{\Gamma(2k)}
			\frac{r^{2k-1/2}\,e^{\ii dr^2/2b}
				}{b^{2k}(-\ii a/b)^{k+\ii\mu}}
	\,	{}_1\!F_{\!1}\left(	{k{+}\ii\mu\atop2k};
				\frac{-\ii r^2}{2ab} \right).
							 \label{hypp-lct1}
\end{gather}
Needless to say, the joint phases of $-\ii a/b$ are needed so as not
fall into multivaluation problems.
This result, reported in~\cite{Basu-KBW}, was calculated following
the general method to f\/ind Mellin transforms of hypergeometric
functions due to Majumdar and Basu~\cite{Majumdar-Basu}.

Now it is only necessary to perform the Mellin transform of~\eqref{hypp-lct1}, to obtain
\begin{gather}
 {}^{2}\!D^k_{\mu,\mu'}\matricita abcd
	:=\left({}^2\Phi^k_\mu,
			\LCT\matricita abcd\,{}^2\Phi^k_{\mu'}\right)
								\nonumber\\ 
\phantom{{}^{2}\!D^k_{\mu,\mu'}\matricita abcd}{}
= e^{-\ii\pi k} 2^{\ii(\mu'-\mu)}
		\frac{\Gamma(k{-}\ii\mu) \Gamma(k{+}\ii\mu')
				}{2\pi\,\Gamma(2k)}  \nonumber\\
\phantom{{}^{2}\!D^k_{\mu,\mu'}\matricita abcd =}{}
 \times b^{-2k}\left(\frac{-\ii d}{b}\right)^{-k+\ii\mu}
			\left(\frac{-\ii a}{b}\right)^{-k-\ii\mu'}
		\,{}_2\!F_{\!1}\left(	{k{-}\ii\mu,\ k{+}\ii\mu'\atop2k};\,
					\frac{1}{ad} \right).
								\label{DHH-minmin2}
\end{gather}
One should note that the complex power functions are evaluated along
the imaginary axis, in the principal sheet where the cut is chosen
along the negative real half-axis.

Let us call this the ${\cal D}_k$-hyperbolic basis (not to confuse it
later on with hyperbolic LCTs), because it lies the ${\cal D}^+_k$
representation of \Lie{Sp($2,\real$)}. The corresponding LCT
with the integral kernel~\eqref{DHH-minmin2} that
transform functions ${\bf f}=\{f(\mu)\}_{\mu\in\real}\in {\cal
L}^2(\real)$ unitarily, are
\begin{gather*}
	{\bf f}_\ssr{M} \equiv \LCT_\ssr{M}:{\bf f}
					= {}^2{\bf D}^k {\bf f},\qquad
	f_{\!\ssr{M}}(\mu)  \equiv   ({\bf f}_\ssr{M})(\mu)
		= \int_{-\infty}^\infty \dd \mu'\,
			{}^2\!D^k_{\mu,\mu'}({\bf M}) f(\mu').
\end{gather*}

Since $J^\gamma_1$ and $J^\gamma_2$ are related by similarity
 through ${\cal C}\frac1{\sqrt{2}}\matricita1{-1}11$ (the
square root of the Fourier transform), we obtain an equivalent
integral LCT,
\begin{gather}
	{}^{1}\!D^k_{\mu,\mu'}\matdos abcd
	={}^{2}\!D^k_{\mu,\mu'}\onehalf
		\matdos{\phantom{-}a{+}b{+}c{+}d}{{-}a{+}b{-}c{+}d
				}{{-}a{-}b{+}c{+}d}{\phantom{-}a{-}b{-}c{+}d}.
					\label{reposc-basis}
\end{gather}
In this section we have thus shown three faces of LCTs:
the summation kernel~\eqref{D0-2} for vectors
$\{f_n\}_{n\in\ZZ^+_0}$, the `radial' canonical transform integral
kernel~\eqref{DO-minmin2} for functions $\{f(r)\}_{r\in\real^+}$ and its
Hankel transform~\eqref{Jplus-eigenbasis}, and the (apparently unknown)
${\cal D}_k$-hyperbolic transforms~\eqref{DHH-minmin2} and~\eqref{reposc-basis} for functions $\{f(\mu)\}_{\mu\in\real}$. These
various functions are in fact the coordinates of the same abstract
vector $\bf f$ as $f_n={}_0\!\langle k,\,k{+}n|{\bf f}\rangle$,
$f(r)={}_-\!\langle k,\,r|{\bf f}\rangle$, $f(\mu)={}_1\!\langle
k,\,\mu|{\bf f}\rangle$, etc. In this sense, all faces of the
Hilbert space vector $\bf f$ are related, and so are their LCTs.

\section[The ${\cal C}^\varepsilon_s$ canonical transforms]{The $\boldsymbol{{\cal C}^\varepsilon_s}$ canonical transforms}
								\label{sec:four}

Let us now consider the range $\gamma<-\frac14$ of the \Lie{so(2,1)}
generators in \eqref{J2andM}, corresponding to centripetal potentials,
which also stem from the separation of $D=2$ coordinates in two
disjoint patches of hyperbolic coordinates. One patch (indicated by
$\sigma=+1$) is $x=\rho\cosh\zeta$, $y=\rho\sinh\zeta$, and the other
($\sigma=-1$) is $x=\rho\sinh\zeta$, $y=\rho\cosh\zeta$, for
$\rho,\zeta\in\real$. Fourier expansion in $\zeta$ provides
$\partial_\zeta^2\mapsto{-\overline{\zeta}^2}$,
$\overline{\zeta}\in\real$ that yields a centripetal term
in the generators~\eqref{J2andM}. In contrast to radial
coordinates, which follow the subgroup reduction $\Lie{Sp($4,\real$)}
\supset \Lie{SO($2$)}\otimes \Lie{Sp($2,\real$)}$, hyperbolic
coordinates conform to the reduction $\Lie{Sp($4,\real$)} \supset
\Lie{O($1,1$)}\otimes \Lie{Sp($2,\real$)}$, where \Lie{O($1,1$)}
contains the discrete ref\/lections $\matricita0110$ that interchanges
the two values of $\sigma$, and $\matricita100{-1}$ that allows the
reduction by even or odd functions in $\rho$, so the range of this
hyperbolic radius is reduced to $r=|\rho|\in[0,\infty)$ and
$\sigma=\hbox{sign}\,\rho$. The Hilbert spaces of functions
for the range of representations $\gamma<-\frac14$ consists of functions
${}_-\!\langle r,\sigma|f\rangle \equiv f( r,\sigma) \equiv f_\sigma( r)$
whose inner product we can represent with two-vector notation
\begin{gather*}
	{\bf f}( r) \equiv \left({f_{+1}\atop f_{-1}}\right)( r)
		\in{\cal L}^2_2(\real^+),  \qquad 
	({\bf f},{\bf g})  :=  \sum_{\sigma\in\{+1,-1\}}
	\int_0^\infty\dd r\,f_\sigma( r)^* g_\sigma( r),
\end{gather*}
taking care to note that $\sigma$ is an index of the basis
functions $|r,\sigma\rangle_{\!-}$ stemming from the two
hyperbolic coordinate patches.
When $\gamma=-\frac14$, the Bargmann index in~\eqref{Bargmann-index}
is $k=\onehalf$; for $\gamma<-\frac14$, $k$ becomes complex,
\begin{gather*}
	k=\onehalf+\ii s,\quad s\in\real,\qquad
			\kappa=\tsty14+s^2\ge\tsty14
\end{gather*}
and determines these representations to belong to the Bargmann {\it
continuous nonexceptional\/} series ${\cal C}_s^\varepsilon$, where
$\varepsilon\in\{0,\frac12\}$ distinguishes between vector and spinor
(two-fold cover) representations of \Lie{SO($2,1$)}, and in the
\Lie{SO($2$)} reduction the index $m\equiv\varepsilon\
\hbox{mod}\,1$ is unrestricted\footnote{I.M.~Gel'fand and M.A.~Na\u{\i}mark called this the {\it principal\/}
representation series~\cite{Gelfand-Naimark,Naimark}.}.

The (reduced) list of eigenfunctions of the generators $J^\gamma_X$
in \eqref{J2andM} is now (cf.~\eqref{Phiminus}),
	\begin{gather}
	{}^0{\bf\Phi}^{\varepsilon,k}_m(r)  =
		\frac{g_\varepsilon(k)}{\pi\sqrt{r}}
		\vecdos{ (-1)^{m-\varepsilon}
			\sqrt{2 \Gamma(k{-}m) \Gamma(1{-}k{-}m)}
				W_{m,k-1/2}(r^2)
			}{ 2 \Gamma(k{+}m) \Gamma(1{-}k{+}m)
				W_{-m,k-1/2}(r^2)},
								\nonumber\\
\phantom{{}^0{\bf\Phi}^{\varepsilon,k}_m(r)  =}{} m{+}\varepsilon\in\ZZ,\qquad	k=\onehalf+\ii s, \qquad
g_0(k)=\cosh\pi s,\qquad g_{1/2}(k)=\sinh\pi s;
							\nonumber\\
	{}^{-}{\bf\Phi}^{\varepsilon,k}_\rho(r)  =
			\vecdos{\delta(\rho{-}r)}{0} \quad \hbox{for }\rho\ge0,\qquad\hbox{and} \qquad
			\vecdos{0}{\delta(|\rho|{-}r)} \quad \hbox{for }\rho<0;\label{Cseries-Phiminus}
			 \\
	{}^{2}{\bf\Phi}^{\varepsilon,k}_{\tau,\mu}(r)  =
			\frac1{\sqrt{2\pi}}\vecdos1\tau
				r^{-1/2+\ii\mu}, \qquad
				   \mu\in\real,\quad \tau\in\{+1,-1\}; \nonumber
	\end{gather}
where $W_{\lambda,\mu}$ is a Whittaker function. The two components
of the eigenfunction are distinguished by the non-symplectic matrix
$\matricita100{-1}\in\Lie{O($1,1$)}$, which on the algebra generators
\eqref{J2andM} and \eqref{JMinusandM} can be interpreted as $2\times2$
matrix operators
${\bf J}^\gamma_0 = \matricita{J^\gamma_0}00{-J^\gamma_0}$,
${\bf J}^\gamma_1 = \matricita{J^\gamma_1}00{-J^\gamma_1}$,
${\bf J}^\gamma_2 = \matricita{J^\gamma_2}00{J^\gamma_2}$,
${\bf J}^\gamma_\pm = \matricita{J^\gamma_\pm}00{-J^\gamma_\pm}$.

\subsection[Elliptic basis: discrete ${\cal C}_s^\varepsilon$-LCTs]{Elliptic basis: discrete $\boldsymbol{{\cal C}_s^\varepsilon}$-LCTs}

The action of the LCT operator ${\cal
C}_\ssr{M}$ on the discrete eigenfunction basis of ${\bf
J}^\gamma_0$, ${}^0{\bf\Phi}^{\varepsilon,k}_m(r)$, for
$m-\varepsilon\in\ZZ$ will be that of an inf\/inite matrix ${\bf
C}^{\varepsilon,k}_\ssr{M}=\Vert C^{\varepsilon,k}_{\ssr{M};\,m,m'}\Vert$.
The transformation of this basis under \Lie{Sp($2,\real$)} can be
again factorized into the right-ordered product in
\eqref{Iwasawa-type}, so that as in \eqref{LCT-on-Phi0}, we now have
\begin{gather}
	\left(\LCT\matricita abcd\,{}^0\!
			{\bf\Phi}^{\varepsilon,k}_m\right)_\sigma(r)
		=\left(\frac{a{-}\ii b}{a{+}\ii b}\right)^m
		\frac{\exp\left(\ii\sigma r^2\frac{ac{+}bd}{2(a^2{+}b^2)}\right)
				}{(a^2+b^2)^{1/4}}\,
		{}^0\!{\bf\Phi}^{\varepsilon,k}_{m,\sigma}
						\left(\frac{r}{\sqrt{a^2{+}b^2}}\right),
				\label{Ccont-act0}
\end{gather}
where the component sign $\sigma$ only appears in the Gaussian
exponent. This relation shows that, in the same way as harmonic
oscillator functions reproduce under the LCTs~\eqref{integral-C1}, and
Bessel functions under radial ${\cal D}_k$ LCTs~\eqref{DO-minmin1}, the
Whittaker functions in~\eqref{Ccont-act0} do likewise under the ${\cal
C}_s^\varepsilon$ continuous-series LCTs.

But the following step of f\/inding the LCT matrix elements as in~\eqref{DoC00}, results in a sum of integrals of two Whittaker
functions and an oscillating Gaussian, which the authors
\cite{Basu-KBW} could not solve. As in
\eqref{D0-2} this result was obtained before with the traditional
$m$-shift operators by Bargmann \cite{Bargmann-1947},
\begin{gather*}
 {}^0\!C^{\varepsilon,k}_{m,m'}\matricita abcd
	:=\left({}^0\!{\bf\Phi}^{\varepsilon,k}_m,\,
		\LCT\matricita abcd\,{}^0\!{\bf\Phi}^{\varepsilon,k}_{m'}\right)
	=\sum_\sigma\int_0^\infty\dd r\, \cdots \\ 
	\hphantom{{}^0\!C^{\varepsilon,k}_{m,m'}\matricita abcd}{}
\overset{\text{for} \ m\ge m'}{=} \ \frac{2^{2m'}}{m'!}
		\sqrt{\frac{\Gamma(k{+}m) \Gamma(1{-}k{+}m)
				  }{\Gamma(k{+}m) \Gamma(1{-}k{+}m)}}
	\frac{[(a{-}d){+}\ii(b{+}c)]^{m-m'}}{[(a{+}d){+}\ii(b{-}c)]^{m+m'}}
					\nonumber\\
	\hphantom{{}^0\!C^{\varepsilon,k}_{m,m'}\matricita abcd
\overset{\text{for} \ m\ge m'}{=} }{} \
 \times\,{}_2\!F_{\!1}\left(
			{k{-}m',\ 1{-}k{-}m'\atop 1{+}m{-}m'};
				-\tsty14(a^2{+}b^2{+}c^2{+}d^2{-}2)
					\right), \\ 
	\hphantom{{}^0\!C^{\varepsilon,k}_{m,m'}\matricita abcd}{}
\overset{\text{for} \ m\le m'}{=} \ (-1)^{m'-m}\frac{2^{2m}}{m!}
		\sqrt{\frac{\Gamma(k{+}m') \Gamma(1{-}k{+}m')
				  }{\Gamma(k{+}m) \Gamma(1{-}k{+}m)}}
	\frac{[(a{-}d){-}\ii(b{+}c)]^{m'-m}}{[(a{+}d){+}\ii(b{-}c)]^{m'+m}}
					\nonumber\\
\hphantom{{}^0\!C^{\varepsilon,k}_{m,m'}\matricita abcd
\overset{\text{for} \ m\le m'}{=}}{} \
 \times\, {}_2\!F_{\!1}\left(
			{k{-}m,\ 1{-}k{-}m\atop 1{+}m'{-}m};
				-\tsty14(a^2{+}b^2{+}c^2{+}d^2{-}2)
					\right). 
\end{gather*}
As in \eqref{CT-0}, but now for inf\/inite vectors of discrete components,
${\bf f}=\{f_m\}_{m-\varepsilon\in\ZZ}$, the LCT is
	\begin{gather*}
	{\bf f}_\ssr{M} \equiv \LCT_\ssr{M}:{\bf f}
					= {}^0{\bf C}^{\varepsilon,k}_\ssr{M} {\bf f},\qquad
	f_{\!\ssr{M}; m}  \equiv   ({\bf f}_\ssr{M})_m
		= \sum_{m'=-\infty}^\infty
			{}^0\!C^{\varepsilon,k}_{m,m'}({\bf M}) f_{m'},
\end{gather*}
and is unitary in the Hilbert space of sequences $\ell^2(\ZZ)$.

\subsection[Parabolic basis: radial ${\cal C}_s^\varepsilon$-LCTs]{Parabolic basis: radial $\boldsymbol{{\cal C}_s^\varepsilon}$-LCTs}

The action of the canonical transform operators $\LCT_\ssr{M}$ on
functions ${}_-\!\langle\sigma,r|f\rangle=f_\sigma(r) \in{\cal
L}^2_2(\real^+)$ will be represented in the continuous series ${\cal
C}_s^\varepsilon$ by a $2\times2$ matrix of integral kernels
	\begin{gather}
	{\bf f}_\ssr{M}(r) \equiv
		\left(\LCT_\ssr{M}:\vectorcito{f_{+1}}{f_{-1}}\right)(r)
					= \int_0^\infty \dd r'\,
		{}^-{\bf C}^{\varepsilon,k}_\ssr{M}(r,r') {\bf f}(r'),\nonumber\\
	{}^-{\bf C}^{\varepsilon,k}_\ssr{M}(r,r') =
	\matdos{{}^-C^{\varepsilon,k}_{\ssr{M}; +1,+1}(r,r')
			}{{}^-C^{\varepsilon,k}_{\ssr{M}; +1,-1}(r,r')
			}{{}^-C^{\varepsilon,k}_{\ssr{M}; -1,+1}(r,r')
			}{{}^-C^{\varepsilon,k}_{\ssr{M}; -1,-1}(r,r')}.
  \label{Ccont-0}
\end{gather}
Here we resort to the reduction $\Lie{Sp($4,\real$)} \supset
\Lie{O($1,1$)}\otimes \Lie{Sp($2,\real$)}$ as presented in~\cite{CT-IV}, with the precisions made in \cite[equations~(2.15)]{Basu-KBW} regarding ranges and phases. For the
non-exceptional continuous series of representations ${\cal
C}_s^\varepsilon$, where $\kappa=k(1{-}k)\ge\frac14$ and
	\begin{alignat}{5}
	 & \varepsilon=0: \quad && h_{\varepsilon}=1,\quad && k{-}\onehalf=\ii s, \quad && s\ge0, &\nonumber\\
	& \varepsilon=\onehalf: \quad && h_{\varepsilon}=-1,\quad &&
			k{-}\onehalf=\ii s,\quad && s>0, &
						\label{cont-exc}
\end{alignat}
where $g_\varepsilon(k)$ is given in \eqref{Cseries-Phiminus}.

The integral kernel elements of ${\bf C}^{\varepsilon,k}_\ssr{M}$ are
then expressed as
\begin{gather}
	({}^-{\bf C}^{\varepsilon,k}_\ssr{M})_{\sigma,\sigma'}(r,r')
		 =  G_{\ssr{M}; \sigma,\sigma'}(r,r')
			H^{\varepsilon,k}_{\sigma,\sigma'}(-rr'/b),
					\label{CGH}\\
	G_{\ssr{M}; \sigma,\sigma'}(r,r')
		 :=  \frac{\sqrt{rr'}}{2\pi\,|b|}
		\exp\left(\ii\frac{d\sigma r^2+a\sigma' r^{\prime\,2}}{2b}\right),
					\label{CGHGG}\\
	H^{\varepsilon,k}_{+1,+1}(\zeta)
		 := \ii\pi\left(e^{-\pi s} H^{(1)}_{2\ii s}(\zeta{+}\ii0^+)
		 -h_{\!\varepsilon} e^{\pi s} H^{(2)}_{2\ii s}(\zeta{-}\ii0^+)\right)
					\nonumber\\
\hphantom{H^{\varepsilon,k}_{+1,+1}(\zeta)}{}
		 =  h_{\!\varepsilon} H^{\varepsilon,k}_{-1,-1}(\zeta)
			= h_{\!\varepsilon} H^{\varepsilon,k}_{+1,+1}(-\zeta)
			=H^{\varepsilon,1-k}_{+1,+1}(\zeta),
					\label{hnhhufids}\\
	H^{\varepsilon,k}_{+1,-1}(\zeta)
		 := 4(-\hbox{sign}\,\zeta) g_\varepsilon(k)\,K_{2\ii s}(|\zeta|)
					\nonumber\\
\hphantom{H^{\varepsilon,k}_{+1,-1}(\zeta)}{}		
=  h_{\!\varepsilon} H^{\varepsilon,k}_{-1,+1}(\zeta)
			= h_{\!\varepsilon} H^{\varepsilon,k}_{+1,-1}(-\zeta)
			= h_{\!\varepsilon} H^{\varepsilon,1-k}_{+1,-1}(\zeta),
					\label{hnhh}
\end{gather}
where $H^{(1)}_\nu$ and $H^{(2)}_\nu$ are the Hankel functions
of the f\/irst and second kind valued above and below the branch cut,
$K_\nu$ is the Macdonald function, while~$g_\varepsilon(k)$ and~$h_\varepsilon(k)$ are given in~\eqref{cont-exc}.

The $2\times2$ kernel LCT in \eqref{Ccont-0},
$C^{\varepsilon,k}_{\ssr{M};\,\sigma,\sigma'}(r,r')$, $r,r'\in\real^+$,
can be written also as
$C^{\varepsilon,k}_{\ssr{M}}(\rho,\rho')$, with $\rho:=\sigma
r\in\real$. In that case, the kernel ${\cal C}_s^\varepsilon$
has been expressed~\cite{Basu-KBW} in terms of conf\/luent hypergeometric
functions,
\begin{gather}
	C^{\varepsilon,k}_\ssr{M}(\rho,\rho')
	 =  \frac{(\hbox{sign}\,b)^{2\varepsilon}
		h_\varepsilon^{(1+{\rm sign}\,\rho')/2} g_\varepsilon(k)
			}{\pi |b|}  		\exp\left(\ii\frac{d\sigma\rho^2
			-2\eta \rho\rho'+a\sigma'\rho^{\prime\,2}}{2b}\right)
			\sqrt{\rho\rho'} \nonumber\\
\hphantom{C^{\varepsilon,k}_\ssr{M}(\rho,\rho')=}{}
\times\left\{ \left[\Gamma(1{-}2k)
		\left|\frac{\rho\rho'}{2b}\right|^{2k-1}
			{}_1\!F_{\!1}\left({2k{-}\onehalf\atop 4k{-}1};\,
				\frac{2\ii\rho\rho'}{\eta b}\right)\right]   + [k\leftrightarrow1{-}k]\right\},
				\label{Cinrho}
\end{gather}
where $\sigma=\hbox{sign}\,\rho$, $\sigma'=\hbox{sign}\,\rho'$;
$\eta=1$ for $\sigma=\sigma'$ and
$\eta=-\ii$ for  $\sigma\neq\sigma'$. In particular, for $b=0$,
\begin{gather*}
	\lim_{b\to0}{}^-C^{\varepsilon,k}_\ssr{M}(\rho,\rho')	
		=\frac{(\hbox{sign}\,a)^{2k}}{\sqrt{|a|}}
			\exp\left(\ii\,\hbox{sign}\,\rho\frac{c\rho^2}{2a}\right)
				\delta(\rho'-\rho/|a|).
\end{gather*}
We are not addressing the Bargmann {\it exceptional\/} representation
series, for which in \eqref{cont-exc} one has $k=\onehalf+s$, with
$0<s<\onehalf$ overlapping the exceptional interval, and for which
\eqref{CGH}--\eqref{hnhh} appears to be a valid integral transform.
Treatment of the exceptional continuous series can be seen in~\cite{Basu-Batta}.

\subsection[Hyperbolic basis and their ${\cal C}_s^\varepsilon$-LCTs]{Hyperbolic basis and their $\boldsymbol{{\cal C}_s^\varepsilon}$-LCTs}

The eigenfunctions ${}^{2}{\bf\Phi}^{\varepsilon,k}_{\tau,\mu}(r)$ of
$J^\gamma_2$ in the continuous series representations form a
generalized basis for a Hilbert space ${\cal L}^2_2(\real)$, and
are given in~\eqref{Cseries-Phiminus} with $\tau\in\{+1,-1\}$,
$\mu\in\real$ and $r\in\real^+$. They can provide an LCT form with a
$2\times2$ matrix  of integral kernels that are the double Mellin
transforms of the LCT kernel in the position basis given by
\eqref{CGH}--\eqref{hnhh} or~\eqref{Cinrho}. Using the technique of Basu
and Majumdar on the latter, \cite{Basu-KBW} reports
\begin{gather*}
 {}^2\!C^{\varepsilon,k}_{\tau,\mu;\tau',\mu'}\matricita abcd
	:=\left({}^2\!\Phi^{\varepsilon,k}_{\tau,\mu},
		\LCT\matricita abcd\,{}^2\!\Phi^{\varepsilon,k}_{\tau',\mu'}\right)
						\\ 
\hphantom{{}^2\!C^{\varepsilon,k}_{\tau,\mu;\tau',\mu'}\matricita abcd}{}
 =  \frac{(-\hbox{sign}\,b)^{2\varepsilon}g_\varepsilon(k)}{2\pi}
		\left[\left(\alpha_k+\frac{\tau\tau'h_\varepsilon}{\alpha_k}
			+\tau'\beta_k +\frac{\tau h_\varepsilon}{\beta_k}\right)
			T_k		\right.	\nonumber\\
\left.\hphantom{{}^2\!C^{\varepsilon,k}_{\tau,\mu;\tau',\mu'}\matricita abcd =}{}
	 + \left(h_\varepsilon\alpha_{1-k}+ \frac{\tau\tau'}{\alpha_{1-k}}
			+\tau'\beta_{1-k}+\frac{\tau h_\varepsilon}{\beta_{1-k}}
			\right)T_{1-k}\right], 
\end{gather*}
where
\begin{gather*}
	T_k :=  \frac{\Gamma(1-2k)\,\Gamma(k-\ii\mu)\,\Gamma(k+\ii\mu')
			}{|a|^{k-\ii\mu'} |b|^{\ii(\mu-\mu')} |d|^{k-\ii\mu}}\,
			{}_2\!F_{\!1}\left(
			{k{-}\ii\mu,\ k{+}\ii\mu'\atop 2k};\,\frac1{ad}	\right),
			\\ 
	\alpha_k := \exp\big(\ii\onehalf\pi[(k+\ii\mu')\,\hbox{sign}\,ab
						+(k-\ii\mu)\,\hbox{sign}\,bd]\big), \\ 
	\beta_k := \exp\big(\ii\onehalf\pi[-(k+\ii\mu')\,\hbox{sign}\,ab
						+(k-\ii\mu)\,\hbox{sign}\,bd]\big), 
\end{gather*}
and where $g_\varepsilon(k)$ and $h_\varepsilon$ are given in~\eqref{cont-exc}.

Note that the two components of this ${\cal L}^2_2(\real)$ are
distinct from the two components of the Hilbert space ${\cal
L}^2_2(\real^+)$ in the parabolic case of the previous subsection. In
this hyperbolic ${\cal C}_s^\varepsilon$ representation, LCTs are
maps of two-component functions by the integral
	\begin{gather*}
	{\bf f}_\ssr{M} \equiv \LCT_\ssr{M}:{\bf f}
					= {}^2{\bf C}^{\varepsilon,k} {\bf f},\nonumber\\
	f_{\!\ssr{M}; \tau}(\mu)
	 \equiv ({\bf f}_\ssr{M})_\tau(\mu)
		= \sum_{\tau'\in\{-1,1\}} \int_{-\infty}^\infty \dd\mu'\,
			{}^2C^{\varepsilon,k}_{\tau,\mu;\tau'\mu'}({\bf M})
					 f_{\tau'}(\mu')
\end{gather*}
and is unitary in that Hilbert space.

\section{Concluding remarks}   \label{sec:five}

The recompilation of the six forms that LCTs have in the elliptic,
parabolic and hyperbolic subgroup bases of \Lie{Sp($2,\real$)} in the
${\cal D}_k$ and ${\cal C}^\varepsilon_s$ representation series, has
been made for the purpose of placing the better-known LCT forms,
i.e., the `linear'~\eqref{integral-C1}, the `radial'~\eqref{the-radial},
and the (lesser known) `hyperbolic' \eqref{CGH}--\eqref{hnhh}, in the
general context of group representation theory. We still have to
justify the appellative of {\it canonical\/} though, because it is a
term associated with the preservation of area elements in classical
phase space, the conservation of energy in paraxial wave optics, and
of uncertainty in quantum mechanics.

Classically, the $2\times2$ symplectic matrix $\MM=\matricita abcd$
in \eqref{Mabcd} acts on the phase-space coordinates, written
$(r,\,p)^\trans$ as if they were a 2-vector. The correspondence~\eqref{sp2R-so21}
carries its action as a $(2{+}1)$-Lorentz transformation of the
3-vector of phase space functions
\begin{gather}
	(\xi_0,\xi_1,\xi_2)^\trans :=
	\left(\tsty14(p^2{+}\gamma/r^2{+}r^2),
		 \tsty14(p^2{+}\gamma/r^2{-}r^2), \tsty12 rp\right)^\trans,
			\label{so21-3vec}
\end{gather}
which close under Poisson brackets into the same algebra, with
$\ii[\cdot,\cdot]\mapsto\Poissbra{\cdot}{\cdot}$,
as the operators $\{J^\gamma_i\}_{i=0}^2$ in \eqref{com-rel-so21},
\begin{gather*}
	\Poissbra{\xi_1}{\xi_2}=-\xi_0,\qquad
	\Poissbra{\xi_2}{\xi_0}=\xi_1,\qquad
	\Poissbra{\xi_0}{\xi_1}=\xi_2.
\end{gather*}
We may consider these relations as the Berezin brackets def\/ining the Lorentz
algebra. The square length of the vector \eqref{so21-3vec} is
$\xi^2_1+\xi^2_2-\xi^2_0=-\tsty14\gamma$ [instead of
$-\tsty14\gamma+\tsty3{16}$ in the operator case~\eqref{Bargmann-index}], and this surface is conserved under Lorentz
transformations. For centrifugal $\gamma>0$ or centripetal $\gamma<0$
these surfaces are respectively a two-sheeted or one-sheeted
hyperboloid, and a cone for $\gamma=0$; these surfaces are also
symplectic manifolds. Since classically
$r=\pm\sqrt{2(\xi_0{-}\xi_1)}$ and $p=2\xi_2/r$, a linear
transformation in the $\xi$'s is nonlinear in $(r,p)^\trans$,
Moshinsky termed these {\it nonlinear\/} canonical transformations~\cite{Mello-Moshinsky, MM-THS-KBW}.

We count the \Lie{Sp($2,\real$)} canonical transforms collected here
as dif\/ferent faces of LCTs acting on dif\/ferent function Hilbert
spaces: $\ell^2(\ZZ_0^+)$ and $\ell^2(\ZZ)$ in the elliptic basis of
${\cal D}_k^+$ and ${\cal C}_s^\varepsilon$ representations
respectively; similarly, ${\cal L}^2(\real^+)$ and ${\cal
L}^2_2(\real^+)$ in the parabolic basis, and ${\cal L}^2(\real)$ and
${\cal L}^2_2(\real)$ in the hyperbolic basis. Of course, the best
known ones are those in the eigenbasis of $J^\gamma_-=\onehalf r^2$
understood as the position observable, which include the `radial'
LCTs in the~${\cal D}_k^+$, particularly the `linear' transform
kernels written by Collins \cite{Collins} and Moshinsky et al.~\cite{Moshinsky-Quesne-osc}, which are genuinely f\/it to describe the
signals or wavefunctions that traverse a paraxial optical or quantum
mechanical system. For the same parabolic subgroup, the `hyperbolic'
LCTs in the ${\cal C}_s^\varepsilon$ representations have found no
proper application, save their use in~\cite{Basu-KBW} to f\/ind
all \Lie{SL($2,\real$)} group representations in all subgroup and
mixed bases reported there. Widely used in other contexts, the
Bargmann results of 1947 \cite{Bargmann-1947} that provide the matrix
representations of the Lorentz group, uses the {\it elliptic\/}
subgroup to provide the row indices. This has been used in~\cite{ANVW00} to propose a covariant discretization for
axis-symmetric $D=2$ systems, where the limit of the discrete to the
continuous radial model is pointedly addressed. Among the six LCT
forms presented above, similar limits from elliptic or hyperbolic to
parabolic subgroup bases should occur in all representations.
Representations referred to the hyperbolic subgroup were investigated
by Mukunda and Radhakrishnan~\cite{Mukunda-Radhakrishnan} but have
also failed to be associated with some model of physical system.

Uncertainty relations are preserved under all LCTs, but they stem
from \Lie{so($2,1$)}~-- not from the Heisenberg--Weyl algebra~-- and
involve the coordinate $\onehalf r^2>0$. Recall that the mean of a~self-adjoint operator $J$ in a wavef\/ield $\psi$ is
$\overline{\jmath}_\psi:=(\psi,J\psi)$, with the inner product
appropriate to its Hilbert space; its dispersion is then
$\Delta_\psi(J):= \Vert (J{-}\overline{\jmath}_\psi)\psi\Vert^2$.
Note also that from the commutator between $J_-=\onehalf r^2$ and its
(Fourier) ${\cal C}\matricita0{-1}10$ transform operator $J_+$, is
$[J_-,J_+]=2\ii J_2$, and it follows that
\begin{gather*}
	\Delta_\psi(J_-) \Delta_\psi(J_+)
		=\Delta_\psi(J_-) \Delta_{\widetilde\psi}(J_-)
		\ge\tsty14|(\psi,J_2\psi)|^2,
\end{gather*}
where $\widetilde\psi={\cal C}\matricita0{-1}10\psi$. In the special
case of the original LCT~\eqref{integral-C1} identif\/ied by~\eqref{osc-rep-134}, this can be written in the form of a
Robertson uncertainty relation $\langle r^2\rangle_\psi
\langle p^2\rangle_\psi\ge\frac14\langle (rp{+}pr)\rangle_\psi^2$
\cite{Robertson}.

We have not addressed in any depth the exceptional interval $0<k<1$,
i.e., $-\frac14\le\gamma<\frac34$ characterizing {\it weak\/}
centripetal and centrifugal potentials, which Bargmann treated with a~nonlocal measure \cite{Bargmann-1947,Basu-Batta}. Also there, the
discrete and exceptional series overlap, and its generators~$J^\gamma_X$ have one-parameter families of self-adjoint extensions
\cite{Fortschritte}, with the result that their spectra are generally
not equally spaced, save for the {\it Friedricks\/} extension in the
${\cal D}_k$ representation series. The various limits to
the point $k=\onehalf$ are particularly troublesome~\cite{Basu-KBW}.

Perhaps most important is the computer use of LCTs to model physical
systems, pursued since the 1990s, in a quest that is running into
hundreds of references \cite{Harper,Sheridan5,Sheridan4, Pei3} where
f\/inite data sets or pixellated images are subject to LCT matrices of
$N\times N$. Of course, we know that noncompact groups such as
\Lie{Sp($2,\real$)} cannot have f\/inite-dimensional unitary
representations. So, approximations have to be made, either through
sampling functions and kernels, or using the f\/inite-dimensional $k<0$
representations of the ${\cal D}_k$ series mentioned in Section~\ref{sec:two}; in both strategies the matrices will not be unitary,
but only in the second will they faithfully represent the LCT group
\Lie{Sp($2,\real$)}.  Perhaps the best approximation strategy is to
use a truncated elliptic basis $\{ {}^0\Phi^k_m(r)\}_{m=k}^{k+N-1}$
of functions that are radial oscillator modes; the matrices would by
neither unitary nor would faithfully represent the group; but if the
signal contains mostly low energy modes it could be fairly
approximated by the f\/irst $N$ modes, and with an inherent
$N\to\infty$ limit to a representation that is both unitary and
faithful.

\subsection*{Acknowledgements}

The Symposium on Superintegrability, Exact Solvability and Special
Functions (Cuernavaca, 20--24 February 2012) was supported by the
Centro Internacional de Ciencias AC, Fondo ``Alfonso N\'apoles
G\'andara'', Instituto de Ciencias F\'isicas, Instituto de
Matem\'aticas, and Intercambio Acad\'emico of the Coordinaci\'on de
la Investigaci\'on Cien\-t\'i\-f\/ica, Universidad Nacional Aut\'onoma
de M\'exico.  This work was supported by the {\it \'Optica
Matem\'atica\/} projects PAPIIT-UNAM 101011 and SEP-CONACyT 79899.

\pdfbookmark[1]{References}{ref}
\LastPageEnding

\end{document}